\documentstyle[12pt]{article}

\begin{document}
\pagestyle{plain}

\newcommand{\be}{\begin{equation}}
\newcommand{\ee}{\end{equation}}
\newcommand{\bea}{\begin{eqnarray}}
\newcommand{\eea}{\end{eqnarray}}
\newcommand{\psib}{\bar{\psi}}

\title{Chern-Simons Term and Charged Vortices in
 Abelian and Nonabelian Gauge Theories}

\author { Avinash Khare\\
{Institute of Physics, Sachivalaya Marg,}\\
{Bhubaneswar 751 005, India.}}

\maketitle

\begin{abstract}
In this article we review some of the recent advances regarding the
charged vortex solutions in abelian and nonabelian gauge theories with
Chern-Simons (CS)
term in two space dimensions. Since these nontrivial results are essentially
because of the CS term, hence, we first discuss in some detail the various
properties of the CS term in two space dimensions. In particular,
 it is pointed out that this parity (P) and time reversal (T) violating but
gauge
 invariant term when added to the Maxwell Lagrangian gives a massive gauge
quanta and yet the theory is still gauge invariant. Further, the vacuum of such
a theory shows the magneto-electric effect. Besides, we show that the CS term
can also
be generated by spontaneous symmetry breaking as well as by radiative
corrections.
A detailed discussion about Coleman-Hill theorem is also given which aserts
that
the parity-odd piece of the vacuum polarization tensor at zero momentum
transfer
is unaffected by two and multi-loop effects. Topological quantization of the
coefficient
of the CS term in nonabelian gauge theories is also elaborated in some detail.

One of the dramatic effect of the CS term is that the vortices of the abelian
(as well as nonabelian) Higgs model now acquire finite quantized charge and
angular momentum. The various properties of these vortices are discussed at
 length with special emphasis on some of the recent developments including the
discovery of the self-dual charged vortex solutions.

\end{abstract}

\section{INTRODUCTION}

In 1957, Abrikosov \cite{ab} wrote down the vortex solutions in Ginzburg-Landau
(GL)
theory which is the mean-field theory of superconductivity. Subsequently, these
vortices were experimently observed in type-II superconductors. In 1973,
Nielsen
and Olesen \cite{no} rediscovered these solutions in the context of the abelian
Higgs
model which is essentially a relativistic generalization of the GL theory.
These people
were looking for string like objects in field theory. It turns out that these
vortex solutions have finite energy per unit length (i.e. finite energy in 2+1
dimensions as the vortex dynamics is essentially confined to the x-y plane),
quantized flux  but are electrically neutral and have zero angular momentum.

In 1975, Julia and Zee \cite{jz} showed that the SO(3) Georgi-Glashow model
which admits
`t Hooft-Polyakov monopole solution also admits its charged generalization i.e.
the dyon solution with finite energy and finite, nonzero charge. It was then
natural
to enquire if the abelian Higgs model which admits neutral vortex solutions
with
finite energy (in 2+1 dimensions), also admits charged vortex solutions with
finite charge and energy in 2+1 dimensions. Julia and Zee showed in the
appendix
of the same paper \cite{jz} that the answer to the question is no. More than
ten years later
Samir Paul(then my Ph.D. student) and myself showed \cite{pk} that the
Julia-Zee
result can be bypassed by adding the CS term \cite{ss,djt} to the abelian Higgs
model in 2+1
dimensions. In particular, we showed that the abelian Higgs model with CS term
in
2+1 dimensions admits charged vortex (soliton to be more precise)
 solutions of finite
energy, nonzero finite charge and flux. As an extra bonus, one found that these
vortices have nonzero angular momentum which is in general fractional. This
strongly suggested that these objects could infact be charged anyons \cite{aka}
i.e. the
objects which are neither bosons or fermions but which obey statistics which is
interpolating between the two. Subsequently, Frohlich and Marchetti \cite{fm}
have shown using axiomatic field theory that these objects are indeed charged
anyons.

There is one question that has remained unanswered though i.e. can one overcome
Julia-Zee objection \cite{jz} in 3+1 dimension itself? Recently we have
answered the question in the
affermative. In particular we \cite{gk} have been able to construct self-dual
topological as well as nontopological charged vortex solutions of finite energy
per unit length in a generalized abelian Higgs model with a dielectric function
and a neutral
scalar field.
The interesting point is that in this case the Bogomol'nyi bound on energy
per unit length is obtained as a linear combination of the magnetic flux and
the
electric charge per unit length.

By now our work on the CS vortices has been extended in several directions.
Most of the developments
till 1988 have been well summarized in my earlier review article on this
subject
\cite{akf}. Howevere, several new advances have taken place since that time and
one purpose of this article is to discuss some of those developments. Since the
key
role in this game is played by the CS term, it is only proper that at first I
discuss the
various properties of this term.

The plan of the paper is as follows: In Sec. II, the role of the CS term is
discussed in the context of both abelian and nonabelian
gauge theories. The key point to note is that whereas the Chern-Pontryagin
term is topological but has no dynamics (being a total divergence), the CS term
is topological and also contributes to equations of motion and hence has
nontrivial
dynamics. In particular, it is pointed out that in 2+1 dimensions,
because of this term, one has at the same time a
massive gauge field and a gauge invariant action.
Quantum electrodynamics with CS term has some interesting and
unusual properties which are discussed here. For example, the vacuum
polarization
tensor has an extra piece which is odd in both P and T . Various properties of
this P and T-odd piece including Coleman-Hill (CH) theorem \cite{ch} and
magneto-electric
effect \cite{kp} are discussed at length. In particular, it is emphasised that
contrary to
the claim of CH, not only fermions but any P and T violating interaction
including
even scalar \cite{kmp} or vector \cite{hpr} particles can give nonzero
contribution to the P and T-odd
part of vacuum polarization tensor at zero momentum transfer. We also point out
that the
theorem is also valid in case the gauge symmetry is spontaneously broken and
the theorem is stated in terms of effective action rather than vacuum
polarization
tensor \cite{kmp}.
In the nonabelian case, it turns out that the theory is not well defined unless
the coefficient of the CS term is
quantized \cite{djt}. Various issues in this regard are discussed at length.
In particular, it is pointed out that the tree level quantization continues
to remain valid  at one \cite{pr} and higher loops \cite{rr} in the case of
pure gauge theories.
Further, it is also valid at one loop in the presence of matter fields.
Finally, it is argued that the quantization of the coefficient of the CS term
is
also valid at one loop for an unbroken nonabelian gauge group in a theory where
larger
nonabelian gauge symmetry is spontaneously broken to that of the smaller gauge
group
\cite{kmpp}.

We also
point out that atleast in the nonabelian case, adding the CS term to the action
is not really a luxary
$-$in a sence one is forced to add it since even if one does not add it at the
tree
level, it is automatically induced by radiative corrections \cite{re}. Finally,
it is pointed
out that a la gauge field mass term the abelian CS term can also be generated
by spontaneous
symmetry breaking \cite{pk1}. This is possible because in 2+1 dimensions, even
for scalar particles
one can introduce a Pauli type nonminimal interaction term. Some other
implications of this
term are also discussed in Sec. II.

In Sec. III we discuss one of the most dramatic consequence of the CS term in
2+1
dimensions i.e. the existance of charged evortex solutions \cite{pk}. Only
salient features
of this solution are discussed here since the details are already contained in
my
previous review article \cite{akf}. In Sec. IV, we discuss some of the recent
developments
in this field. In particular, we discuss the self-dual charged vortex solutions
in both relativistic \cite{jw} and nonrelativistic \cite{jp} theories as well
as
semi-local charged vortex solutions \cite{aks}.

\section{Some Important Properties of the Chern-Simons Term}

It is well known that there is a gauge singlet (popularly known as axial)
anomaly in any even dimension $2n$ so that the divergence of the gauge
singlet axial current is proportional to the corresponding Chern-Pontryagin
density $P_{2n}$ in that dimension which in turn can always be written as a
total
divergence i.e.
\be \label{2.1}
P_{2n} =\partial_{\mu} \Lambda^{\mu} ; \ \mu = 0,1,2,...... n-1
\ee
The object $\Lambda^{\mu}$, for a particular value of $\mu$ (say $\mu_0$)
naturally lives
in $(n-1)$ dimensions (excluding that particular dimension $\mu_0$) and is
known
as the Chern-Simons density. For example the gauge singlet anomaly in 3+1
dimensional quantum electrodynamics is given by
\bea \label{2.2}
\partial^{\mu} j_{\mu}^5 = {e^2\over 2\pi}\epsilon_{\mu\nu\lambda\sigma}
F^{\mu\nu} F^{\lambda\sigma}\\
 =  {e^2\over \pi}\partial^{\mu}
(\epsilon_{\mu\nu\lambda\sigma}A^{\nu} F^{\lambda\sigma})
\eea
so that the abelian CS term in 2+1 dimensions is given by
\be \label{2.3}
\Lambda^3 = {e^2\over \pi} \epsilon_{3\nu\lambda\sigma} A^{\nu}
F^{\lambda\sigma}
\ee
which clearly lives in 2+1 dimensions. In the nonabelian case, the CS term has
an extra piece i.e.
\be \label{2.4}
\Lambda^3 \propto \epsilon^{\nu\lambda\sigma} Tr (F_{\nu\lambda}
A_{\sigma}-{2\over 3} A_{\nu}A_{\lambda}A_{\sigma})
\ee
where $A_{\mu}$ and $F_{{\mu}{\nu}}$ are matrices
\be \label{2.5}
A_{\mu} = gT^a A_{\mu}^a\ ; \ F_{\mu\nu} = gT^a F_{\mu\nu}^a
=\partial_{\mu}A_{\nu}-\partial_{\nu}A_{\mu}
+[A_{\mu},A_{\nu}]
\ee
Here we have used the representation matrices of the group (for SU(2): $T^a =
\tau ^a/2i$)
\be \label{2.6}
[T^a , T^b] = f^{abc} T^c .
\ee

Let us first discuss the properties of the abelian CS term as given by eq.
(\ref{2.3}).

{\bf P and T Violation:} It is easily seen that the abelian (as well as the
nonabelian) CS term is not invariant
under P and T seperately even though it is invariant under PT as well as charge
conjugation C. It is worth noting here that in 2+1 dimensions, even the fermion
mass term ${m \overline\psi \psi}$ is odd under P as well as T and infact this
is the
underlying reason as to why CS term can be generated in perturbation theory by
integrating over fermions in a massive fermionic theory \cite{re}.

{\bf CS Term as Gauge Field Mass Term:} Let us consider electrodynamics in the
presence of the CS term
\be \label{2.7}
{\cal L} = - {1\over 4} F_{\mu\nu} F^{\mu\nu} + {\mu\over 4}
\epsilon^{\mu\nu\lambda} F_{\mu\nu} A_{\lambda}
\ee
where $F_{{\mu}{\nu}}$ = ${{\partial_{\mu} A_{\nu}} - {\partial_{\nu}
A_{\mu}}}$. Note that
the corresponding equation of motion
\be \label{2.8}
\partial_{\mu} F^{\mu\nu} + {\mu\over 2} \epsilon^{\nu\alpha\beta}
F_{\alpha\beta} = 0
\ee
as well as the action are invariant under $U(1)$ gauge transformation
\be \label{2.9}
 A_{\mu} \rightarrow A_{\mu} - {1\over 2} \partial_{\mu} \Lambda
\ee
The field eq. (\ref{2.8}) can also be written as
\be \label{2.10}
(g^{\mu\nu} +{1\over \mu} \epsilon^{\mu\nu\alpha}\partial_{\alpha})
^*F_{\nu} = 0
\ee
where $^* F_{\nu}$ is the dual field strength which is a vector in three
dimensions
\be \label{2.11}
^*F_{\nu} = {1\over 2} \epsilon_{\nu\alpha\beta} F^{\alpha\beta} ; \
F_{\mu\nu} = \epsilon_{\mu\nu\alpha} {^*F^{\alpha}}
\ee
On operating by $(g_{\beta\mu} - \epsilon_{\beta\mu\delta} \partial^{\delta}
/\mu)$
 to eq. (\ref{2.10}) we have
\be \label{2.12}
( \Box + \mu^2 ) ^*F_{\beta} = 0
\ee
which clearly shows that the gauge field excitations are massive. This
remarkable
property of having a gauge invariant mass term for the gauge field in the
action itself is very special to 2+1 dimensions. In all other dimensions
one has to take recourse to the Higgs mechanism (or one could have dynamical
symmetry
breaking as in the 1+1 dimensional Schwinger model). It is woth noting here
though that unlike
in other dimensions, in this case, both massless and the CS-mass photon has
only
one degree of freedom. Further, whereas the massless photon in 2+1 dimensions
has spin zero, the CS-mass photon has spin 1 (-1) if $\mu > (<) 0$. Note that
because
of the CS term, one has necessarily a P and T violating theory.
On the otherhand,
the normal massive photon has two degrees of freedom and both spins $\pm 1$ are
present as they should be in a parity conserving theory.

{\bf Coleman-Hill Theorem:} It turns out that because of the P and T violating
but gauge invariant CS term, the most general form for the vacuum polarization
tensor (consistent with Lorentz and gauge invariance) is more general than in
other dimensions i.e.
\be \label{2.13}
\Pi_{\mu\nu} (k) = (k^2
g_{\mu\nu}-k_{\mu}k_{\nu})\Pi_1(k^2)-i\epsilon_{\mu\nu\lambda}
k^{\lambda}\Pi_2 (k^2)
\ee
Notice that the second term on the r.h.s. is P and T odd. It is clear that any
P and T violating interaction will contribute to $\Pi_2(k^2)$. For example the
fermion mass term which in 2+1 dimensions break both P and T, does contribute
to
$\Pi_2(k^2)$ at one loop level. Remarkably enough, it was discovered that at
two
loops, though, there is no contribution to $\Pi_2(0)$ \cite{sk}. Inspired by
this result, Coleman and Hill \cite{ch} have infact proved under very general
conditions that $\Pi_2(0)$ receives no contributions from two and higher loops
in any gauge and Lorentz invariant theory including particles of spin one or
less.
In particular, they have emphasized that their result is valid even for
nonrenormalizable interactions in the presence of gauge and Lorentz invariant
regularizations. These authors also claimed that at one loop the only
contribution
to $\Pi_2(0)$ can come from fermion loop. This is however not true. In
particular,
there is no reason why P and T violating interactions involving spin 0 or 1
particles
should not contribute to $\Pi_2(0)$ at one loop. Indeed, Hagen et al.
\cite{hpr} as well as
we \cite{kmp} have shown that nonrenormalizable spin one and spin zero
interactions respectively do contribute to $\Pi_2(0)$ at one loop.

It might be added here that apart from these situations which were overlooked
by
Coleman-Hill, there are other situations where the initial assumptions of the
theorem are not satisfied and where $\Pi_2(0)$ does get further radiative
corrections.
One such situation is if there are massless particles present in which case
infrared
divergences spoil the proof of the theorem \cite{ssw}. Another case is if
Lorentz
or gauge invariance is not satisfied, a situation found in the nonabelian case.
A third case is that of spontaneously broken scalar electrodynamics \cite{ks}
where the term quadratic in the gauge field explicitly violates one of the
assumption of Coleman and Hill. However, even in this case we have recently
shown
\cite{kmp} that if the theorem is formulated in terms of effective action
rather
than vacuum polarization tensor then the coefficient of the CS term in the
effective
action does not receive a radiative correction at one loop.

{\bf Magnetoelectric Effect:} There are many crystals in nature like cromium
oxide which show this effect i.e. they get magnetized in an electric field and
electrically polarized in a magnetic field \cite{ll}. It is well known that
this effect
depends upon having a T-assymmetric medium. In this case the usual relation
between $\vec D$ and $\vec E$ as well as between $\vec H$ and $\vec B$ is
modified to
\be \label{2.14}
D_i =\chi_{ij}^{(e)} E_j + \chi_{ij}^{(em)} B_j
\ee
\be \label{2.15}
H_i =\chi_{ij}^{(m)} B_j + \chi_{ij}^{(me)} E_j
\ee
Since CS term violates P and T, it is natural to ask if the vacuum of the 2+1
dimensional QED with CS term also shows the magnetoelectric effect or not. We
have shown \cite{kp} that indeed the vacuum in such a theory does show this
effect and both $\chi_i^{(em)}$ and $\chi_i^{(me)}$ are proportional to $k_i
\Pi_2(k^2)$.

{\bf CS Term by Spontaneous Symmetry Breaking:} We have seen above that the CS
term provides mass to the gauge field. Now usually the gauge field mass is
generated by
spontaneous symmetry breaking hence it is worth enquiring if the CS term can
also
be generated by this mechanism. The answer to the question turns out to be yes.
This is because as has been shown by us \cite{pk1}, unlike other dimensions,
2+1
dimensions offer more general possibility of a covariant derivative. For
example,
 it is possible to introduce a Pauli type nonminimal coupling for even scalar
particles. In particular, notice that
\be \label{2.16}
D_{\mu}\phi = (\partial_{\mu}+ie
A_{\mu}+ig\epsilon_{\mu\nu\lambda}\partial^{\nu}A^{\lambda})\phi
\ee
also behaves like a covariant derivative since the nonminimal term by itself
is gauge covariant. As a result one now finds that the matter field kinetic
energy
term ${1\over 2} (D^{\mu}\phi)^{*} (D_{\mu}\phi)$ has a piece $eg |\phi|^2
\epsilon_{\mu
\nu\lambda}(\partial^{\mu} A^{\nu}) A^{\lambda}$. Thus if $\phi$ acquires a
nonzero
vacuum expectation value then the (abelian) CS term is generated. Clearly a
similar mechanism should also work for the nonabelian case, but technically it
is
a tougher problem since one also has to generate the nonlinear triple gluon
coupling term. So far as I know, till today it is an open problem.

The resulting nonminimal theory has been shown to have some very interesting
properties
in case the magnetic coupling constant g (called critical magnetic moment)
acquires
a special value
\cite{st,gw}. In particular, this theory gives rise to an effective action
which is
renormalizable at one loop \cite{ku} and has similar properties to the one
describing ideal
anyons upto an additive contact term.
It is worth adding here that this nonminimal magnetic coupling can be induced
by radiative corrections even if it is not present at the tree level \cite{ko}.

Yet another remarkable property of the CS term is that in this case the Lorentz
invariance of the action automatically follows from gauge invariance. In
particular,
whereas for the Maxwell case the most general gauge invariant Lagrangian is
\be \label{2.17}
{\cal L} =\vec E^2 + a \vec B^2
\ee
it is only the demand of Lorentz invariance which fixes a to be -1. On the
other
hand, in the CS case the demand of gauge invariance automatically fixes the
form
of the CS term.

Let us now discuss some properties which are unique to the nonabelian CS term.

{\bf Quantization of the CS Mass:} In the nonabelian gauge theory with CS term
in 2+1 dimensions one finds that the gauge field is again massive and that the
theory is well
defined only if this mass is infact quantized. The gauge field Lagrangian is
\be \label{2.18}
{\cal L} = {1\over 2g^2} Tr (F^{\mu\nu}F_{\mu\nu}) - {\mu\over
2g^2}\epsilon^{\mu\nu\alpha} Tr (F_{\mu\nu}A_{\alpha}-{2\over 3}
A_{\mu}A_{\nu}A_{\alpha})
\ee
where $A_{\mu}$ and $F_{\mu\nu}$ are matrices as defined by eq. (\ref{2.5})
The field equation which follows from here
\be \label{2.19}
D_{\mu} F^{\mu\nu} + {\mu\over 2} \epsilon^{\nu\alpha\beta}
F_{\alpha\beta} = 0
\ee
where
\be \label{2.20}
D_{\mu} = \partial_{\mu} + [ A_{\mu},  ]
\ee
is gauge covariant. As in the abelian case it immediately follows that the
gauge field has mass $\mu$. Now notice that the action $I_{CS}$ = $\int d^{3}x
\cal L_{CS}$
 even though invariant under small gauge transformations, is not invariant
under
homotopically-nontrivial gauge transformations \cite{djt}. In particular, if
the
gauge group $G$ is such that
\be \label{2.21}
\Pi_3 (G) = Z
\ee
where $Z$ is the group of integers (note in particular that eq. (\ref{2.21}) is
true for any gauge group of which SU(2) is a subgroup), then under these so
called
large gauge transformations the action transforms as
\be \label{2.22}
I_{cs}\longrightarrow I_{cs} + {8\pi^2\mu\over g^2} m
\ee
where m is an integer. Now in the path integral formulation, the action itself
may or may not be gauge invariant but what is required is that the exponential
of the action should atleast be gauge invariant. We thus conclude that the
nonabelian gauge theory with CS term does not make sence in 2+1 dimensions
unless the CS mass is quantized in units of $g^{2}/4 \pi$ i.e.
\be \label{2.23}
{8\pi^2\mu\over g^2} = 2n\pi \ or \ \mu = {g^2\over 4\pi} n\ , n = 0,
\pm 1, \pm 2,...
\ee
Please note that in 2+1 dimensions the gauge coupling g is not dimensionless
but rather has dimension of $(mass)^{1\over 2}$.

{\bf Parity Anomaly:} Someone might wonder as to why is one considering models
with CS term in the first place since afterall this term violates both P and T.
 The answer to that is (atleast in nonabelian gauge theories) even if one does
not add CS term to the action at the tree
level, it is still generated by the radiative corrections$-$the effect due to
the so called
parity anomaly \cite{re}. In particular, even though the action
\be \label{2.24}
I [A_{\mu},\psi ] = \int d^3 x[{1\over 2g^2} Tr
F_{\mu\nu}F^{\mu\nu} +i\overline\psi \gamma_{\mu} D^{\mu} \psi ]
\ee
is invariant under both gauge transformations and P and T, the effective action
$I_{eff}[A]$ obtained by integrating out the fermionic degrees of freedom must
violate one of the two symmetries in the case of odd number of massless
fermions.
In other words, there is no regularization which can simultaniously maintain
the invariance of $I_{eff}$ under parity as well as under large gauge
transformations.
In particular, under large gauge transformations with winding number n,
$I_{eff}$
transforms as
\be \label{2.25}
I_{eff} [A] \longrightarrow I_{eff} [A ] \pm n\pi\, \ n = 0,\pm 1,\pm 2,....
\ee
Since gauge invariance is usually regarded as more important than parity, one
can
maintain gauge invariance at the cost of parity  by adding the P and T-odd CS
term
in the action (or by regulating the theory in such a way that the CS term is
automatically produced
$-$say by using Pauli-Villers regularization). In this way one finds that
because
of the parity anomaly the CS term is induced by radiative corrections even if
it is
absent at the tree level. This is very similar to the way the Chern-Pontryagin
term is induced in even dimensions due to the gauge singlet (chiral) anomaly.
Note however that whereas the anomaly in even dimensions is in the divergence
of
the current, in 2+1 case there is no anomaly in the divergence of the current.
The
anomaly is in the current itself in that there is a piece with wrong parity.
Also notice that the parity anomaly is only in the nonabelian gauge
theories and not in the abelian U(1) case since in this case there are no large
gauge transformations
in the first place!

Finally, since the CS action depends only on $\epsilon_{\mu \nu\lambda}$ tensor
and not on the metric
$g_{\mu\nu}$ hence the gauge field action with only nonabelian (or even
abelian) CS term is
an example of topological field theory \cite{wi}.

\section{Charged Vortex Solutions}

Let us consider the abelian Higgs model with CS term as given by the Lagrangian
\be \label{3.1}
{\cal L} = - {1\over 4} F_{\mu\nu}F^{\mu\nu} + {1\over 2}(\partial_{\mu} -
ieA_{\mu})\phi^* (\partial^{\mu}+ie A^{\mu})\phi - C_4
(\mid\phi^2\mid - {C_2\over 2C_4})^2 + {\mu\over 4}
\epsilon_{\mu\nu\lambda} A^{\lambda} F^{\mu\nu}
\ee
Following the neutral vortex case \cite{no}, let us consider the following
n-vortex ansatz
\be \label{3.2}
\vec A (\vec\varphi,t) = - \hat e_0 C_0 {g(r)-n\over r},\
\phi(\vec\varphi,t) = C_0 e^{in\theta}f(r),\ A_0(\vec\varphi,t) = C_0 h(r)
\ee
where
\be \label{3.3}
\rho = {r\over eC_0} , \ C_0 =\sqrt{{C_2\over 2C_4}}
\ee
We have rescaled the lengths and the fields so that one can work in terms
of the dimensionless variables. It turns out that the dynamics essentially
depends on two dimensionless parameters $\delta$ and $\lambda$ defined by
\be \label{3.4}
\lambda =\sqrt{(8 C_4/e^2)} ; \ \delta = \mu/eC_0
\ee
The field equations which follow from here are \cite{pk}
\be \label{3.5}
g'' (r) - {1\over r} g' (r) -gf^2 = r\delta h'(r)
\ee
\be \label{3.6}
h'' (r) + {1\over r} h' (r) -hf^2 = {\delta\over r} g'(r)
\ee
\be \label{3.7}
f''(r) + {1\over r} f'(r) - {g^2f\over r^2} +{\lambda^2\over 2}
f(1-f^2) = - h^2f
\ee
while the field energy can be shown to be \cite{pk}
\be \label{3.8}
E_n = \pi C^2_0 \int^{\infty}_0 r dr \bigg [ {1\over r^2} ({dg\over
dr})^2+ ({df\over dr})^2+({dh\over dr})^2 +h^2 f^2 + {g^2f^2\over
r^2} +{\lambda^2\over 4} (1-f^2)^2 \bigg ]
\ee
Several remarks are in order at this stage.

(i) As expected, in the limit $A_{o} = 0$ (i.e. h = 0) and $\mu = 0$ (i.e.
$\delta = 0$) the field equations reduce to those of the neutral vortex case
\cite{no}. From the Gauss law eq. (\ref{3.6}) it also follows that if
$\mu$ is nonzero then $A_o$ must also be nonzero.

(ii) The boundary conditions for finite energy solutions are
\be \label{3.9}
\lim_{r\rightarrow\infty} : f(r) = 1, \ h(r) = 0, \ g(r) = 0
\ee
\be \label{3.10}
\lim_{r\rightarrow 0} : f(r) = 0, \ h(r) = \beta, \ g(r) = n
\ee
with $\beta$ being an arbitrary number.

(iii) A la neutral vortex case, in this case also one has flux quantization
since the boundary conditions are again such that $\Pi_{1}(U(1)) = Z$
\be \label{3.11}
\Phi\equiv \int Bd^2 x = - {2\pi\over e}\int^{\infty}_0 rdr {1\over
r} {dg\over dr} = {2\pi\over e} n
\ee
{}From the Gauss law eq. (\ref{3.6}) it then follows that these vortices also
have nonzero and finite quantized charge \cite{pk}
\be \label{3.12}
Q\equiv \int e^2 hf^2 d^2 x = \mu \Phi = {2\pi\mu\over e} n
\ee
thereby bypassing the Julia-Zee objection.
As far as I am aware off, this is
probably the first time that the quantization of the Noether charge has
followed from purely topological considerations. In a way the relation
(\ref{3.12})
can be looked upon as the 2+1 analog of the Witten effect \cite{wit}. One can
also
compute the magnetic moment of these vortices and show that whereas the neutral
vortices have it equal to the flux $\Phi$ (= $2\pi n/e$), the
charged ones have an extra piece
\be
K_{z} = \int (\vec r\times \vec j)_z d^2 x =  {2\pi n\over e}
+{\delta\over e}\int^{\infty}_0 h(r) d^2 r
\ee

(iv) As an extra bonus, one also finds that unlike the neutral vortices, the
charged vortices have nonzero angular momentum which is ingeneral fractional
and quantized in units of $Q/2e$ i.e.
\be \label{3.13}
J \equiv \int d^2 x \epsilon^{ij} x_i T_{oj} = - {nQ\over 2e}
\ee
This strongly suggests that the charged vortices could infact be charged anyons
\cite{aka}. By explicit construction of a quantum one vortex operator, Frohlich
and Marchetti
\cite{fm} have rigorously shown that this is indeed the case. Thus charged
vortices
provide us with a relativistic field theory model of extended charged anyons.

So far, no analytic solution has been obtained to the field eqs. (\ref{3.5}) to
(\ref{3.7}). However, it is easily seen that for large r, the asymptotic values
of the gauge and Higgs field are reached exponentially fast
\be \label{3.14}
g(r) = \alpha \sqrt r e^{-m_v r}
\ee
\be \label{3.15}
h(r) = {\alpha\over \sqrt r} e^{-m_v r}
\ee
\be \label{3.16}
 f(r) = 1+\beta e^{-\lambda r}
\ee
where $\alpha, \beta$ are dimensional constants while
\be \label{3.17}
m_v = \sqrt{{\mu^2\over 4}+e^2  C^2_0} - {\mu\over 2} .
\ee
Naively, another solution with
\be \label{3.18}
m_v = \sqrt{{\mu^2\over 4}+e^2  C^2_0} + {\mu\over 2} ,
\ee
is also possible but as has been shown in \cite{in}, such a solution does not
exist
for all $r$. It is worth noting here that because of the P and T violating CS
term, the gauge field, after the Higgs mechanism, propagates two modes each
with one degree of freedom and with $J$ = 1(-1) if $\mu >(<) 0$ \cite{pr,pk2}.
It is easily seen that the field equations are invariant under $r
\longrightarrow -r$ so
that the behaviour of the fields around r=0 is given by
\be \label{3.19}
g(r) = n + \alpha r^2 + 0 (r^4)
\ee
\be \label{3.20}
 h(r) = \beta + \alpha \delta {r^2\over 2} + 0(r^4)
\ee
\be \label{3.21}
f(r) = \alpha_2 r^{\mid n\mid }+ 0 (r^{\mid n\mid +2})
\ee
The qualitative behaviour of the charged vortex is as follows: the magnetic
field
B decreases monotonically from its nonzero value at the core of the vortex
(r=0) to zero at $r =\infty$ with penetration length $1/m_v$ while the scalar
field
increases from zero at origin to its vacuum value at infinity with coherence
length $1/m_s$. Finally, the electric field  $E_{\rho}$ vanishes at both r=0
and
$r =\infty$ reaching the maximum in between at some
finite r. It is worth pointing out that as in the Hall effect, for the charged
vortex solutions too $\vec E$ (= $E_{\rho}$) is at right angles to $\vec j$
(=$j_{\phi}$) and both in turn are at right angles to B. It is also worth
pointing
out here that according to the presently accepted explanation,
the quasi-particles responsible for the fractionally quantized Hall effect are
the charged vortices.

\section{Recent Advances}

After our discovery of the charged vortex solutions in 1986 \cite{pk}, in last
ten years this work has been extended in several directions.
I shall only briefly discuss the developments till 1988 which are already
contained
in my previous review article on this subject \cite{akf} while will discuss in
detail
some of the latter developments.

{\bf Charged Vortex-Vortex Interaction:} Perhaps the most interesting question
is if we can directly observe the charged
vortices in some condensed matter system. In this context recall that the
neutral
vortices with one unit of vorticity have been seen in type-II superconductors.
However, none has been seen in type-I superconductors. This can be understood
from the fact that in the neutral case, whereas the vortex-vortex interaction
is attractive in type-I region ($\lambda <1$), it is repulsive in type-II
region ($\lambda >1$) \cite{jr}. It is thus of great interest to study
charged vortex-vortex interaction and see as to when is it repulsive. This
has been done by us \cite{jkkp} using perturbation theory in CS mass as well as
by a variational calculation. For example, when CS mass is small, one can
expand the charged
vortex fields in terms of the neutral vortex fields plus corrections in powers
of CS mass $\delta$. In particular, it has been shown that to $O(\delta^2)$,
the
charged n-vortex fields are given by \cite{jkkp,in}
\be \label{4.1}
 g(\lambda,\delta) = g_0(\lambda) + {r^2\delta^2\over 8}
g_0(\lambda) + 0 (\delta^4)
\ee
\be \label{4.2}
 h(\lambda,\delta ) = {\delta\over 2} g_0(\lambda) + 0 (\delta^3)
\ee
\be \label{4.3}
f(\lambda,\delta ) = f_0(\lambda) + 0 (\delta^4)
\ee
where $g_o$ and $f_o$ are the solutions to the corresponding neutral vortex
solutions in the absence of the CS term. On substituting the solution as given
by eqs. (\ref{4.1}) to (\ref{4.3}) in the expression for the field energy as
given by eq. (\ref{3.8}) one can show that \cite{jkkp}
\be \label{4.4}
E_n^{cha}(\lambda,\delta) = E_n^{neu}(\lambda) +
{n^2\delta^2\over 4} + 0 (\delta^4)
\ee
It is worth noting that the $O(\delta^2)$ correction is positive, proportional
to $n^2$ and independent of $\lambda$. From this equation it immediately
follows
that
\be \label{4.5}
E_n^{cha}(\lambda,\delta) - n \ E_1^{cha}(\lambda,\delta) = E_n^{neu}(\lambda)
-n E_1^{neu}(\lambda) + {(n^2-n)\over 4} \delta^2 + 0 (\delta^4)
\ee
so that the charged vortex-vortex interaction is more repulsive than the
corresponding
neutral case with the extra repulsion coming from the electric field of the
charged vortex. We have also performed a variational calculation (which is more
reliable than the perturbative calculation for larger values of $\delta$) and
find that even in this case the same picture continues
to hold good. For example, for $\delta$ = 0.5 the charged vortex-vortex
interaction is repulsive even for $\lambda > 0.45$ (note that in the neutral
case the vortex-vortex interaction is repulsive only when $\lambda > 1$).

A word of caution is in order here. Our analysis is only valid in the case of
superimposed vortices. The problem of charged vortex-vortex interaction when
the
vortices are seperated by distance d, is still an open unsolved problem.

{\bf Pure CS vortices:} Can one obtain charged vortex solutions in abelian
Higgs
model with pure CS term (i.e. no Maxwell kinetic energy term)? This question is
particularly sensible in the condensed matter context since in the long wave
length
limit the CS term dominates over the Maxwell term. In this context note that
the
Higgs mechanism is operative even in the absence of the Maxwell term and even
in this case one obtains both massive gauge and scalar fields \cite{dy}. The
question of the charged vortices in the abelian Higgs model with pure CS term
was
addressed by us \cite{jk} and we showed that the charged vortex solutions are
indeed possible in this case. Their properties are almost same as those of the
Maxwell-CS
charged vortices except that the magnetic field is now zero at the core of the
vortex and is concentrated in a ring surrounding the vortex core \cite{jk}. It
is worth noting here that in the absence of the Maxwell term, the gauge field
eqs.
(\ref{3.5}) and (\ref{3.6}) are already of first order
\be \label{4.6}
 -g f^2 = r\delta h' (r)
\ee
\be \label{4.7}
 h f^2 = {\delta\over r} g'(r)
\ee
while eq. (\ref{3.7}) remains unaltered and is still a coupled second order
equation. The obvious interesting question is if one can also write it as a
coupled
first order equation so that a la Bogomol'nyi \cite{bo} one could obtain
self-dual
charged vortex solutions. This question was raised by us \cite{jk} but we were
unable to obtain the first order equation. It was left to two other groups
\cite{jw} to make this important
breakthrough. They showed that in addition to dropping the Maxwell term one
also has
to replace the usual $\phi^4$ potential with the following $\phi^6$-type
potential
\be \label{4.8}
V(\mid\phi\mid) = {e^4\over 8\mu^2}\mid\phi^2\mid (\mid\phi^2\mid - v^2)^2
\ee
so as to obtain the self-dual equations. In that case, the self-dual equations
turn out to be
\be \label{4.9}
 f' = \pm {1\over r} fg
\ee
\be \label{4.10}
 B \equiv - {1\over r} {dg\over dr} = \pm {1\over 2} f^2 (1-f^2)
\ee
while
\be \label{4.11}
 h = \mp {1\over 2} (1-f^2)
\ee
One can infact decouple the eqs. (\ref{4.9}) and (\ref{4.10}) and obtain the
following uncoupled second order equation in f
\be \label{4.12}
 f''(r) + {1\over r} f'(r) - {f'^2\over f} + {1\over 2} f^3 (1-f^2)
= 0
\ee
These self-dual equations are quite similar to those of the corresponding
neutral
case (at $\lambda$ = 1) which are given by
\be \label{4.13}
f' = \pm {1\over r} gf
\ee
\be \label{4.14}
 B\equiv - {1\over r} {dg\over dr} = \pm {1\over 2} (1-f^2)
\ee
and hence the uncoupled second order equation in that case is
\be \label{4.15}
f'' (r) + {1\over r} f'(r) - {f'^2\over f} +{1\over 2} f (1-f^2) = 0
\ee
Further, whereas the Lagrangian for the self-dual neutral vortex case is the
bosonic part of a N = 1 supersymmetric theory \cite{df}, the Lagrangian for the
charged self-dual vortex case is the bosonic part of a N = 2 supersymmetric
theory \cite{llw}.

Before we discuss the solutions to the self-dual eqs. (\ref{4.9}) and
(\ref{4.10})
it may be worthwhile to point out that the usual $\phi^4$ potential and the
$\phi^6$
potential as given by eq. (\ref{4.8}) represent very different physical
situations
\cite{bk}. Whereas the $\phi^4$ potential in eq. (\ref{3.1}) corresponds to the
case
of second order phase transition with $T < T_c^{II}$, the $\phi^6$ potential
given above
corresponds to the case of first order transition with $T = T_c^{I}$. One way
to understand as to why $\phi^6$-type
potential is required for the CS vortices while $\phi^4$ potential is required
for
the neutral vortex case is that whereas in four space-time dimensions the
coefficient
of the $\phi^4$-term is dimensionless, it is the coefficient of the
$\phi^6$-term which is
dimensionless in 2+1 dimensions.

It turns out that the self-dual eqs.
(\ref{4.9}) and (\ref{4.10}) admit both topological and nontopological
self-dual charged vortex solutions. Let us first discuss the topological
solutions.

{\bf Topological Self-dual Solutions:} The topological solutions satisfy the
boundary conditions as given by eqs. (\ref{3.9}) and (\ref{3.10}) with $\beta$
= $\mp 1/2$ (for n $> (<) 0$ respectively). The flux, the
charge and the angular momentum of these vortices are as given by eqs.
(\ref{3.11})
to (\ref{3.13}) respectively and the energy is $\pi v^2 |n|$. Infinite number
of sum rules have been derived for these vortices \cite{ak1} (as well as the
neutral self-dual vortices
\cite{ak2}) and using these we have shown that the magnetic moment of the
self-dual
pure CS vortex is $2\pi n(n+1) \mu^2 /{e^3v^2}$. It is worth emphasizing that
the self-dual solutions have been obtained not only with the cylindrical ansatz
but also with an arbitrary ansatz. Further, following the work of Taubes for
the
neutral self-dual vortices \cite{ta}, Wang \cite{wa} has given rigorous
argument for the
existance of self-dual charged vortex solutions even when the vortices are not
superimposed on each other but lie at arbitrary positions in the plane.

{\bf Nontopological Self-dual Solutions:} Since the potential as given by eq.
(\ref{4.8})
has degenerate minima at $\phi = 0$ as well as at $\mid\phi\mid = v$, one finds
that apart from
the topological, one also has nontopological self-dual charged vortex solutions
provided one chooses the following boundary conditions \cite{ak1,jlw}
\be \label{4.16}
\lim_{r\rightarrow\infty} : f(r) = 0 , \ g(r) = \mp \alpha, \
\alpha > 0
\ee
\be \label{4.17}
\lim_{r\rightarrow 0} : f(r) = 0 , \ g(r) = n, \ if \ n \not = 0
\ee
\be \label{4.18}
\lim_{r\rightarrow 0} : f(r) = \eta , \ g(r) = 0 \ if \ n = 0
\ee
where $+ (-)\alpha$ is for $n > (<) 0$.
The flux, the charge and the angular momentum of these vortices can be shown to
be ($n>0$)
\be \label{4.19}
\Phi = {2\pi\over e} (n+\alpha) ; \ Q = \mu\Phi ; \ J = {\pi\mu\over
e^2} (\alpha^2-n^2) ; \ E = \pi v^2 (n+\alpha)
\ee
It is worth pointing out that the finiteness of energy requires that $\alpha >
1$
but otherwise $\alpha$ is arbitrary. However, again in this case infinite
number of sum rules have been derived by us \cite{ak2} using which we have been
able to
show that $\alpha$ must infact satisfy the lower bound $\alpha \ge (n+2)$
\cite{ak4}. Further,
from these sum rules it also follows that the magnetic moment of these
nontopological
vortices is $-2\pi(\alpha + n)(\alpha -n -1)\mu^2 /{e^3v^2}$. As far as I am
aware off,
this is the first instance when both topological and nontopological self-dual
solutions simultaneously exist in a given model. So far as the decay to charged
scalar
mesons is concerned, these nontopological solutions are
at the edge of their stability \cite{jkb}. Finally, Spruck and Yang have
rigorously shown the existance of
self-dual nontopological CS vortices even when they are not superimposed on
each other
but lie at arbitrary positions in the plane \cite{sy}.

{\bf Various Other Self-dual Solutions:} Once the self-dual CS vortices were
discovered in 1990, there has been a flurry of activity and several people have
obtained various other self-dual solutions. While it is clearly impossible to
mention all these developments, we shall try to note atleast some of them. For
example, since $\phi^4$ and $\phi^6$-type models correspond to very different
physical situations, hence it is of interest to enquire if self-dual charged
vortex solutions can also be obtained in the original $\phi^4$ model \cite{pk}
itself. This has been done and Lee et al. \cite{llm} have shown that such
solutions
can also be constructed in the $\phi^4$ model \cite{pk} provided one adds a
neutral scalar field in the model. Further, self-dual solutions have also been
constructed with unusual properties \cite{pg} by essentially multiplying the
Maxwell
and/or the CS term by a dielectric function (which in almost all cases is
assumed to
be a function of the scalar field $\phi$ alone) and also by including the
nonminimal interaction
term. For example, a class of self-dual solutions have
been obtained which are degenerate in energy but have different flux which is
not
quantized.
Further, nonabelian self-dual CS vortices have also been obtained
\cite{leecug}.

{\bf Nonrelativistic Self-dual Vortices:} In an interesting paper, Jackiw and
Pi
\cite{jp} started with the abelian Higgs model with pure CS term and the
$\phi^6$
potential as given by eq. (\ref{4.8}) (which has both topological and
nontopological
self-dual charged vortex solutions) and considered its nonrelativistic limit.
In particular, they showed that to leading order in the velocity of light c,
this model
reduces to
\be \label{4.20}
{\cal L}_{NR} = {\mu\over 4} \epsilon_{\mu\nu\lambda} F^{\mu\nu}
A^{\lambda} + i\psi^* (\partial_t+iA_0)\psi - {1\over 2m} (D_i\psi)^*
(D_i\psi) +{1\over 2mc\mid\mu\mid} (\psi^*\psi)^2
\ee
where $\psi$ represents the particle part of the mode expansion of the scalar
field $\phi$ (the anti-particle
part having been put equal to zero) and m is the mass of the field $\phi$. This
nonrelativistic model can be looked upon either as a classical field theory
or as a second quantized N-body problem with 2-body attractive
$\delta$-function
interaction. These authors were able to obtain self-dual nontopological charged
vortex solutions with zero energy in the above model and showed that these
vortex
solutions are precisely the nonrelativistic limit of the corresponding
nontopological
charged vortex solutions. In fact, the flux, the charge and the angular
momentum
of these objects turn out to be the same as given by eq. (\ref{4.19}) but where
$\alpha = (n+2)$ i.e. the lower bound on $\alpha$ \cite{ak2} is saturated
in the nonrelativistic case. By now, this work has been extended in several
directions.
Mention may be made of the time dependent solutions by Ezawa et al.
\cite{ez} and the nonrelativistic Maxwell-CS vortices \cite{dt}. Further,
nonrelativistic
limit of the nonabelian self-dual CS vortices has also been considered and
interesting
connections with integrable models have been discovered \cite{gdjpt}.

{\bf Interaction Between Self-Dual CS Vortices:} Following the work of Manton
\cite{ma} for the case of monopole and neutral vortices, recently Kim and Min
\cite{km} have considered the slow motion of two well seperated CS vortices
and have shown that the effective Lagrangian (with finite degrees of freedom)
has a statistical interaction term and that this term reflects the anyonic
nature
of the CS vortices. This analysis has recently been extended to the
nonrelativistic
case \cite{hc}.

{\bf Semi-Local Self-Dual CS Vortices:} Recently, semi-local neutral vortex
solutions
have been obtained in an abelian Higgs model with $SU(N)_{global} \otimes
U(1)_{local}$
symmetry. The key point of the argument is that even though topologically
trivial, these solutions are stable under small perturbations due to the
gradient energy term
\cite{hind}.
These semi-local solutions gave some initial hope of finding energatically
stable solutions in the Weinberg-Salam model. However, subsequent analysis has
shown that these semi-local vortex solutions are always unstable for the
realistic
values of the Weinberg angle. Inspired by this work, we have constructed
\cite{aks}
semi-local self-dual CS vortex solutions in an abelian Higgs model with pure
CS term and with $SU(N)_{global} \otimes U(1)_{local}$ symmetry. Most of the
results for the
CS vortices can be easily extended to this case.

{\bf Charged Vortices of Finite Energy Per Unit length in 3+1 Dimensions:}
So far we have shown that the Julia-Zee objection \cite{jz} can be bypassed in
2+1
dimensions by adding CS term to the action. However, the question remains if
one
can also overcome their objection in 3+1 dimensions itself (remember that
their original argument
is infact in 3+1 dimensions)? Recently, we \cite{gk} have been able to do that.
In particular, we showed that if one generalizes the abelian Higgs model by
adding a
dielectric function and a neutral scalar field then one can obtain self-dual
topological as well as nontopological charged vortex solutions with finite
energy
per unit length. In particular, we considered the following generalized abelian
Higgs model
\bea \label{4.21}
{\cal L} = - {1\over 4} G(\mid\phi\mid) F_{\mu\nu} F^{\mu\nu}+ {1\over 2}
\mid (\partial_{\mu}-i eA_{\mu})\phi \mid^2+{1\over
2}G(\mid\phi\mid)\partial_{\mu}N\partial^{\mu}N \nonumber \\
 - {e^2\over
8G(\mid\phi\mid)} (\mid\phi\mid^2 - v^2)^2 - {e^2\over 2} N^2\mid\phi\mid^2
\eea
where $N$ is the neutral scalar field. We showed that if the dielectric
function
$G(\mid\phi\mid)$ is chosen to have the form
\be \label{4.22}
 G(\mid\phi\mid) = {\alpha\over \mid\phi\mid^2}
\ee
then the self-dual equations of the model can essentially be mapped to those of
the purs CS vortices thereby explaining as to why one has both topological as
well as
nontopological charged vortex solutions. However, unlike in that case,
the Bogomol'nyi bound on energy is expressed not only in terms of
the flux but also the vortex charge per unit length.

\pagebreak

\pagebreak

\begin{thebibliography}{99}

\bibitem{ab} A.A. Abrikosov, Sovt. Phys. JEPT {\bf 5} (1957) 1174.
\bibitem{no} H.B. Nielsen and P. Olesen, Nucl. Phys. {\bf B 61} (1973) 45.
\bibitem{jz} B. Julia and A. Zee, Phys. Rev. {\bf D 11} (1975) 2227.
\bibitem{pk} S.K. Paul and A. Khare, Phys. Lett. {\bf B 174} (1986) 420;
{\bf B 182} (1986) E 414.
\bibitem{ss} W. Siegel, Nucl. Phys. {\bf B 156} (1979) 135 ; J. Schonfeld,
Nucl. Phys. {\bf B 185} (1981) 157.
\bibitem{djt} S. Deser, R. Jackiw and S. Templeton,
Ann. of Phys. {\bf 140} (1982) 372.
\bibitem{aka} J.M. Leinaas and J. Myrheim, Nuo. Cim. {\bf B 37} (1977) 1, For a
recent review of this field see A. Khare, Current Sc. (India) {\bf 61} (1991)
826.
\bibitem{fm} J. Frohlich and P.A. Marchetti, Comm. Math. phys. {\bf 121} (1989)
177.
\bibitem{gk} P.K. Ghosh and A. Khare, Bhubaneswar preprint IP/BBSR/94-14,
hep-th 9404015.
\bibitem{akf} A. khare, Forts. der phys. {\bf 38} (1990) 507.
\bibitem{jw} J. Hong, Y. Kim and P.Y. Pac, Phys. Rev. Lett. {\bf 64} (1990)
2330 ;
R. Jackiw and E.J. Weinberg, Phys. rev. Lett. {\bf 64} (1990) 2334.
\bibitem{jp} R. Jackiw and S-Y. Pi, Phys. Rev. Lett. {\bf 64} (1990) 2969 ;
Phys.
Rev. {\bf D 42} (1990) 3500.
\bibitem{aks} A. Khare, Phys. Rev. {\bf D 46} (1992) R 2287.
\bibitem{ch} S. Coleman and B. Hill, Phys. Lett. {\bf B 159} (1985) 184.
\bibitem{kp} A. Khare and T. Pradhan, Phys. Lett. {\bf B 231} (1989) 178.
\bibitem{kmp} A. Khare, R.B. MacKenzie and M.B. Paranjape, Phys. Lett. {\bf B }
(1995) In Press.
\bibitem{hpr} C. R. Hagen, P.K. Panigrahi and S. Ramaswami, Phys. Rev. Lett.
{\bf 61}
(1988) 389.
\bibitem{pr} R.D. Pisarski and S. Rao, Phys. Rev. {\bf D 32} (1985) 2081.
\bibitem{rr} G. Giavarini, C.P. Martin and F. Ruiz Ruiz, Nucl. Phys. {\bf B
381} (1992) 222.
\bibitem{kmpp} A. Khare, R.B. MacKenzie, P.K. Panigrahi and M.B. Paranjape,
Univ.
de Montreal preprint UdeM-LPS-TH-150, hep-th/9306027. This paper had raised
this
interesting question but because of numerical mistakes in computation they
arrived
at wrong conclusion. Recently the correct answer to the issue has been given
by L. Chen, G. Dunne, K. Haller and E. Lim-Lombridas, Univ. of Connecticut
preprint
UCONN-94-8, hep-th 9411062.
\bibitem{re} A.N. Redlich, Phys. Rev. {\bf D 29} (1984) 2366.
\bibitem{pk1} S.K. Paul and A. Khare, Phys. Lett. {\bf 193} (1987) 253.
\bibitem{sk} Y. Kao and M. Suzuki, Phys. Rev. {\bf D 31} (1985) 2137 ;
M. Bernstein and T. Lee, Phys. Rev. {\bf D 32} (1985) 1020.
\bibitem{ssw}  V.P. Spiridonov, JETP Lett. {\bf 52} (1990) 513 ; V.P.
Spiridonov
and F.V. Tkachov, Phys. Lett. {\bf B 260} (1991) 109.
\bibitem{ks} S. yu. Khlebnikov, JETP Lett. {\bf 51} (1990) 81 ; V.P.
spiridonov,
Phys. Lett. {\bf B 247} (1990) 337.
\bibitem{ll} L.D. Landau and E.M. Lifshitz, Electrodynamics of Continuous
Media,
Second Edition, Pergamon Press (1963) ; T.H. O'Dell, The Electrodynamics of
Magneto-Electric Media, North Holland (1970).
\bibitem{st} J. Stern, Phys. Lett. {\bf B 265} (1991) 119.
\bibitem{ku} M.E. Carrington and G. Kunstatter, Phys. Lett. {\bf B 321} (1994)
223.
\bibitem{gw} Y. Georgelin and J.C. Wallet, Int. J. Mod. Phys. {\bf A7} (1992)
1149.
\bibitem{ko}  I.I. Kogan, Phys. Lett. {\bf B 262} (1991) 83.
\bibitem{wi} E. Witten, Comm. Math. Phys. {\bf 121} (1989) 351 ; For the
abelian
case see, M. Bos and V.P. Nair, Phys. Lett. {\bf B223} (1989) 61.
\bibitem{wit} E. Witten, Phys. Lett. {\bf B 86} (1979) 283.
\bibitem{in} V.I. Inozemstsev, Euro. Phys. Lett. {\bf 5} (1988) 113 ; G.
Lozano,
M.V. Manias and F.A. Schaposnik, Phys. Rev. {\bf D 38} (1988) 601.
\bibitem{pk2} S.K. Paul and A. Khare, Phys. Lett. {\bf B 171} (1986) 244.
\bibitem{jr} L. Jacobs and C. Rebbi, Phys. Rev. {\bf B 19} (1979) 4486.
\bibitem{jkkp} L. Jacobs, A. khare, C.N. Kumar and S.K. Paul, Int. J. Mod.
Phys. {\bf A 6}
(1991) 3441.
\bibitem{dy} S. Deser and Z. Yang, Mod. Phys. Lett. {\bf A 4} (1989) 2123.
\bibitem{jk} D.P. Jatkar and A. Khare, Phys. Lett. {\bf B 236} (1990) 283.
\bibitem{bo} E.B. Bogomol'nyi, Sov. J. Nucl. Phys. {\bf 24} (1977) 449.
\bibitem{df} P. di Vecchia and S. Ferrara, Nucl. Phys. {\bf B 130} (1977) 93.
\bibitem{llw} C. Lee, K. Lee and E. Weinberg, Phys. Lett. {\bf 243} (1990) 105.
\bibitem{bk} S.N. Behera and A. Khare, Pramana (J. Phys., India) {\bf 15}
(1980) 245.
\bibitem{ak1} A. Khare, Phys. Lett. {\bf B 255} (1991) 393.
\bibitem{ak2} A. Khare, Phys. Lett. {\bf B 277} (1992) 123.
\bibitem{ta} C. Taubs, Comm. Math. Phys. {\bf 72} (1980) 277.
\bibitem{wa} R. Wang, Comm. Math. Phys. {\bf 137} (1991) 587.
\bibitem{jlw} R. Jackiw, K. Lee and E. Weinberg, Phys. Rev. {\bf D 42} (1990)
3488.
\bibitem{ak4} A. Khare, Phys. Lett. {\bf B 263} (1991) 227.
\bibitem{jkb} D.P. Jatkar and A. Khare, J. Phys. {\bf A 24} (1991) L1001 ; D.
Bazeia,
Phys. Rev. {\bf D 43} (1991) 4074.
\bibitem{sy} J. Spruck and Y. Yang, Comm. math. Phys. {\bf 149} (1992) 361.
\bibitem{llm} C. Lee, K. Lee and H. Min, Phys. Lett. {\bf B 255} (1990) 79.
\bibitem{pg} P.K. Ghosh, Phys. Lett. {\bf B 326} (1993) 264; Phys. Rev. {\bf D
49} (1994) 5458 ;
J. Lee and S. Nam, Phys. Lett. {\bf B 261} (1991) 79 ; M. Torres, Phys. Rev.
{\bf D 49} (1994) R 2295.
\bibitem{leecug} K. Lee, Phys. Rev. Lett. {\bf 66} (1991) 553 ; L.F.
Cugliandolo
et al., Mod. Phys. Lett. {\bf A 6} (1991) 479.
\bibitem{ez} Z. Ezawa, N. Hotta and A. Iwazaki, Phys. Rev. {\bf D 44} (1991)
452.
\bibitem{dt} G. Dunne and C. Trugenberger, Phys. Rev. {\bf D 43} (1991) 1332.
\bibitem{gdjpt} B. Grossman, phys. Rev. Lett. {\bf 65} (1990)  ; G. Dunne, R.
Jackiw,
S-Y. Pi and C. Trugenberger, Phys. Rev. {\bf D43} (1991) 1332.
\bibitem{ma} N. Manton, Phys. Lett. {\bf B 110} (1982) 54 ; Phys. Lett. {\bf B
154}
(1985) 397 ; P.I. Ruback, Nucl. Phys. {\bf B 296} (1988) 669 ; T.M. Samols,
Phys.
Lett. {\bf B 244} (1990) 285.
\bibitem{km} S.K. Kim and H. Min, Phys. Lett. {\bf B 281} (1992) 81.
\bibitem{hc} L. Hua and C. Chou, MIT preprint MIT-CTP No. 2064 (1992).
\bibitem{hind} M. Hindmarsh, Phys. Rev. Lett. {\bf 68} (1991) 1263 ; G.W.
Gibbons,
M.E. Ortiz, F. Ruiz Ruiz and T.M. Samols, Nucl. Phys. {\bf 385} (1992) 127 .

\end{thebibliography}
\end{document}